  \documentstyle[twoside,11pt,draft]{article}
  \def\be{\begin{equation}}
  \def\ee{\end{equation}}
  \def\de{{\rm d}}
  \def\De{{\rm D}}
  
  \title{The Born-Oppenheimer Approach to the Matter-Gravity System and Unitarity}
  \author{C. Bertoni \\
  {\small\it Istituto di Radioastronomia, via Gobetti 101, 40129 Bologna Italy}
  \and F. Finelli, G. Venturi \\
  {\small\it Dipartimento di Fisica, Universit\`a degli Studi di Bologna
  and I.N.F.N.} \\
  {\small\it via Irnerio, 46 -- 40126 Bologna -- Italy}
  }
  
\begin{document}
  \baselineskip 4.0ex
  \begin{titlepage}
  \pagestyle{empty}
  \maketitle
  \begin{abstract}

The Born-Oppenheimer approach to the matter-gravity system is illustrated 
in a simple minisuperspace model and the corrections to quantum field theory 
on a semiclassical background exhibited. Within such a context the unitary 
evolution for matter, in the absence of phenomena such as tunnelling or other 
instabilities, is verified and compared with the results of other approaches. 
Lastly the simplifications associated with the use of adiabatic invariants 
to obtain the solution of the explicitly time dependent evolution equation for 
matter are evidenced.

 \end{abstract}
  
  \end{titlepage}
  \pagestyle{plain}
  \raggedbottom
  \setcounter{page}{1}

  The Born-Oppenheimer (BO) approach has been extensively applied 
  to composite systems, such as molecules, which involve two mass, or time,
  scales \cite{mead}. 
  Such an approach has also been suggested for the matter-gravity
  quantum system \cite{brout} in order to generalize the suggestion that 
  matter follows semiclassical gravity adiabatically \cite{banks} (in the quantum
  mechanical sense). The plausibility of such an approach relies on the
  fact that the mass scale of gravity is the Planck mass which is much
  greater than that of normal matter. Thus one may consider the matter
  variables as the "fast" degrees of freedom whereas the "slow" ones
  are the gravitational variables. This is in analogy with the case of 
  molecules where one considers the intermolecular distance to be the slow
  degree of freedom and the electron coordinates to be the fast ones.
  
  The purpose of this note is to briefly illustrate and compare the diverse
  approaches \cite{kiefer1,kiefer2,kim} 
  with a particular emphasis on a possible violation of unitarity
  in the evolution of the matter system. For this it will be sufficient
  to consider a Friedmann-Robertson-Walker (FRW) line element:
  
  \be
  \de s^2 = - \de t^2 + a^2(t) g_{ij} \de x^i \de x^j =
  a^2(\eta) \left( - \de \eta^2 + g_{ij} \de x^i \de x^j \right)
  \ee
  
  \noindent 
  where $ g_{ij}$ is the metric for a three-space of constant curvature $k$
  (which we shall always take equal to $1$)
  and we have introduced a conformal time $\eta$. We shall consider a
  matter-gravity action given by:
  
  \be
  S = \int \de \eta \left[ - \frac{1}{2} m^2 \dot{a}^2 - m^2 V 
  + L^M(a, \phi) \right]
  \ee
  
  \noindent 
  where a dot denotes a derivative with respect to $\eta$, $m$ is the Planck
  mass, $m^2 V$ the gravitational potential \cite{coscos} 
  and $L^M$ the matter ($\phi$)
  Lagrangian which is allowed to depend on $a$ but not $\dot{a}$. From the
  above one obtains a classical Hamiltonian:
  
  \be
  H = - \frac{\pi_a^2}{2 m^2} + m^2 V(a) + H^M(a,\phi)
  \ee
  
  \noindent where $\pi_a = - m^2 \dot{a}$, $H^M$ is the matter Hamiltonian (which
  does not depend on $\pi_a$) and the classical Hamiltonian constraint
  is $H=0$. We note that the momentum constraint (diffeomorphism
  invariance on a space-like three-surface) is automatically satisfied in this
  minisuperspace model.
   
  On quantising one obtains a Wheeler-DeWitt (WDW) equation \cite{dewitt}:
  
  \be
  \hat{H} \Psi(a,\phi) \equiv \left( \frac{\hbar^2}{2 m^2} 
  \frac{\partial^2}{\partial a^2} + m^2 V(a) + \hat{H}^M \right)
  \Psi(a,\phi) =0
  \label{wdw}
  \ee
  
  \noindent 
  where $\Psi$ is a function of $a$ and $\phi$ and describes both gravity
  and matter. Subsequentely one makes a BO
  decomposition of $\Psi$ as:
  
  \be
  \Psi(a, \phi) = \psi(a) \chi(a,\phi)
  \ee
  
  \noindent 
  where $\chi(a,\phi)$ is not further separable. Coupled equations of motions
  for $\chi$ and $\psi$ may then be obtained by first substituing the above 
  decomposition into eq. (\ref{wdw}) and contracting with $\chi^*$ obtaining
  \cite{brout}:
  
  \begin{eqnarray} \lefteqn{
  \left[ \frac{\hbar^2}{2m^2} \De^2 + m^2 V + \frac
  {\langle \chi | \hat{H}^M | \chi \rangle}{\langle \chi | \chi \rangle}
  \right] \psi = - \frac{\hbar^2}{2 m^2} 
  \frac{\langle \chi | \bar{\De}^2 | \chi \rangle}{\langle \chi | \chi 
  \rangle} \psi = 
  } \nonumber \\ & &
  \frac{\hbar^2}{2 m^2} \frac{1}{\langle \chi | \chi \rangle}
  \left\{ \langle \chi |
  \frac{\stackrel{\leftarrow}{\partial}}{\partial a} 
  \left( 1 - 
  \frac{| \chi \rangle \langle \chi |}
  {\langle \chi |\chi \rangle } \right)
  \frac{\partial}{\partial a} |
  \chi \rangle \right\} \psi
  \label{eqpsi}
  \end{eqnarray}
  
  \noindent where we have introduced covariant derivatives:
  
  \be
  \De \equiv \frac{\partial}{\partial a} + i A \,; \;\;\;\;\;
  \bar{\De} \equiv \frac{\partial}{\partial a} - i A
  \ee
  
  \noindent with \cite{norsta}:
  
  \be
  A \equiv - i \frac{\langle \chi | \frac{\partial}{\partial a} \chi \rangle} 
  {\langle \chi | \chi \rangle} \equiv - i \langle \frac{\partial}{\partial a} 
  \rangle
  \ee
  
  \noindent and a scalar product:
  
  \be
  {\langle \chi | \chi \rangle} \equiv \int \de \phi \chi^*(a,\phi) \chi(a,\phi)
  \ee
  
  \noindent where the integral is over the different matter modes.
  
  On now multiplying eq. (\ref{eqpsi}) by $\chi$ and subtracting it 
  from eq. (\ref{wdw}) one obtains:
  
  \begin{eqnarray} \lefteqn{
  \psi \left( \hat{H}^M - \langle \hat{H}^M \rangle \right) \chi +
  \frac{\hbar^2}{m^2} \left( \De \psi \right) \bar{\De} \chi =
  - \frac{\hbar^2}{2 m^2} \psi
\left( \bar{\De}^2 - \langle \bar{\De}^2 \rangle \right)
  \chi =
  } \nonumber \\ & &
  - \frac{\hbar^2}{2 m^2} \psi \left[ \left( 
  \frac{\partial^2}{\partial a^2} - \langle \frac{\partial^2}{\partial a^2}
  \rangle \right) - 2 \langle \frac{\partial}{\partial a} \rangle 
  \left( 
  \frac{\partial}{\partial a} - \langle \frac{\partial}{\partial a}
  \rangle \right)  \right] \chi
  \label{eqchi}
  \end{eqnarray}
  
  \noindent 
  and we note that the r.h.s. of eqs. (\ref{eqpsi}) and (\ref{eqchi}) are
  related to fluctuations, that is they consist of an operator acting on
  a state minus its expectation value with respect to that state, and in the
  BO (or adiabatic) approximation they are neglected.
  
  In order to better understand our equations it is convenient to consider 
the following form for the gravitational wavefunction:
  
  \be
  \psi  \equiv e^{- i \int^a A \de a'} \tilde{\psi}
  \simeq \frac{1}{N} e^{- i \int^a A \de a' + \frac{i}{\hbar} S^{eff} }
  \label{psitil}
  \ee
  
  \noindent where $N$ (which is related to the Van Vleck determinant
  \cite{banks,kiefer1}) and $S^{eff}$ are real. On neglecting fluctuations 
in eq. (\ref{eqpsi}), 
in the semiclassical limit for gravity $S^{eff}$ will satisfy the following
  gravitational Hamilton-Jacobi equation \cite{clalim}:
  
  \be
  - \frac{1}{2 m^2} \left( \frac{\partial S^{eff}}{\partial a} \right)^2
  + m^2 V + \langle \hat{H}^M \rangle = 0
  \ee
  
  \noindent which includes the back reaction of matter through the 
  average matter Hamiltonian. We note $\frac{\partial S^{eff}}{\partial a} =
  - m^2 \dot{a}$ and in the classical limit also for matter one obtains 
  the classical Einstein equation.
  
  On substituting into eq. (\ref{eqchi}) one obtains:
  
  \begin{eqnarray} \lefteqn{
  \left( \hat{H}^M - i \hbar \frac{\partial}{\partial\eta} \right)
  e^{ - \frac{i}{\hbar} \int^\eta \langle \hat{H}^M \rangle \de \eta' -
  i \int^{a(\eta)} \de a' A} \chi =
  } \nonumber \\ & & 
  - \frac{\hbar^2}{m^2} e^{ - \frac{i}{\hbar} \int^\eta \langle \hat{H}^M 
\rangle \de \eta' -
  i \int^{a(\eta)} \de a' A} \left[ - \frac{\partial \log N}{\partial a}
  \bar{\De} + \frac{1}{2} \left( \bar{\De}^2 - \langle \bar{\De}^2 
\rangle \right)
  \right] \chi
  \label{sch1}
  \end{eqnarray}
  
  \noindent where we have introduced the time derivative through 
$
 i \hbar \frac{\partial}{\partial \eta} \equiv 
- i \frac{\hbar}{m^2} \frac{\partial S^{eff}}
  {\partial a} \frac{\partial }{\partial a} \,.
$
  We note that on neglecting the r.h.s. of eq. (\ref{sch1}),
  corresponding to the semiclassical limit for gravity and the
  adiabatic (or BO) approximation, one obtains the usual evolution equation
  for matter (Schwinger-Tomonaga or Schr\"{o}dinger).
  Thus it is natural to identify the matter wave function in the above as
  \cite{phases}:
  
  \be
  \chi_s \equiv e^{- \frac{i}{\hbar} \int^\eta \langle \hat{H}^M \rangle d\eta'
  - i \int^{a(\eta)} da' A} \chi \equiv
  e^{- \frac{i}{\hbar} \int^\eta \langle \hat{H}^M \rangle d\eta'} \tilde{\chi}
  \label{chisch}
  \ee
  
  \noindent 
  since in the above mentioned limits it becomes the usual Schr\"{o}dinger
  wave function.
  
  One may now search for eventual violations of unitary evolution for
  matter by considering:
  
  \begin{eqnarray} \lefteqn{
  i \hbar \frac{\partial}{\partial \eta} \int \chi_s^* \chi_s \de \phi =
  \int \de \phi \left[ \chi_s^* i \hbar \frac{\partial}{\partial \eta} \chi_s
  - c.c. \right] =
  } \nonumber \\ & &
e^{-i \int^{a(\eta)} da' (A-A^{\dagger})} \int \de \phi \left\{ \chi^*
\left[ \hat{H}^M - \frac{\hbar^2}{m^2}
  \frac{\partial \log N}{\partial a} \bar{\De}
  + \frac{\hbar^2}{2 m^2} \left( \bar{\De}^2 - \langle \bar{\De}^2 \rangle
  \right) \right]\chi - c.c. \right\} =
\nonumber \\ & &
e^{-i \int^{a(\eta)} da' (A-A^\dagger)} 
\left\{ \left[ \langle \chi| \hat{H}^M | \chi \rangle - \frac{\hbar^2}{m^2}
\frac{\partial \log N}{\partial a} \langle \chi| \bar{\De} |\chi \rangle 
\right. \right.
\nonumber \\ & &
\left. \left.
+ \frac{\hbar^2}{2 m^2} \langle \chi |
\left( \bar{\De}^2 - \langle \bar{\De}^2 \rangle
\right) | \chi \rangle \right] - c.c. \right\} = 0 \,,
\label{unit1}
\end{eqnarray}
  
  \noindent 
  where we observe that $\langle \bar{\De} \rangle$, $\langle \left( 
\bar{\De}^2 - \langle \bar{\De}^2 \rangle \right) \rangle$ are zero and
we have assumed $\langle \hat{H}^M \rangle $ is real.  Thus unless one
  considers non Hermitian Hamiltonians (or the presence of tunneling phenomena
  leading to instabilities) there is no violation of unitarity. We note that in
  our approach, wherein we identified the gravitational and matter wave
  functions and equations of motion, one could envisage the presence of 
  instabilities both in matter and gravitation and such that they
  {\em compensate} in the composite system \cite{venturi1}.
  
  Let us now compare our results with those of others. One approach 
\cite{kiefer1}
  consists
  of expressing $\Psi$ as:
  
  \be
  \Psi = e^{\frac{i S}{\hbar}}
  \ee
  
  \noindent and expanding S in powers of $m^2$:
  
  \begin{eqnarray} \lefteqn{
  S =  m^2 S_0 + S_1 + \frac{1}{m^2} S_2 + {\cal O} \left( \frac{1}{m^4} \right)
  } \nonumber \\ & &
  = \left[ m^2 S_0(a) + \frac{1}{m^2} \sigma_2(a) \right] +
  \left[ S_1(a,\phi) + \frac{1}{m^2} \eta_2(a,\phi) \right] + 
{\cal O} \left( \frac{1}{m^4} \right) \,.
  \end{eqnarray}
  
  \noindent 
This allows one to rewrite $\Psi$ in the following factorized form:
  
  \be
  \Psi \approx \left( \frac{1}{N_K} e^{ \frac{i}{\hbar} m^2 S_0 +
  \frac{i}{m^2 \hbar} \sigma_2 } \right) \left( N_K 
  e^{ \frac{i S_1}{\hbar} + \frac{i \eta_2}{m^2 \hbar}} \right) \equiv
  \tilde{\psi}_K \tilde{\chi}_K
  \label{psichikief}
  \ee
  
  \noindent
  where $N_K$ is also related to the Van Vleck determinant.
  We may now tentatively identify the above
  $\tilde{\psi}_K$ and $\tilde{\chi}_K$ with our $\tilde{\psi}$ 
  (eq. (\ref{psitil})) and $\tilde{\chi}$ (eq. (\ref{chisch})) which
  satisfy:
  
  \be
  \left( \frac{\hbar^2}{2 m^2} \frac{\partial^2}{\partial a^2} +
  m^2 V + \langle \hat{H}^M \rangle \right) \tilde{\psi} =
  - \frac{\hbar^2}{2 m^2} \frac{\langle \tilde{\chi} | \frac{\partial^2}
{\partial a^2} | \tilde{\chi} \rangle }
{\langle \tilde{\chi} | \tilde{\chi} \rangle} 
\tilde{\psi}
  \label{hjtilde}
  \ee
  
  \noindent and:
  
  \be
  \left( \hat{H}^M - \langle \hat{H}^M \rangle \right) \tilde{\chi}
  +  \frac{\hbar^2}{m^2} \frac{\partial \ln \tilde{\psi}}{\partial a}
  \frac{\partial \tilde{\chi}}{\partial a} =
  - \frac{\hbar^2}{2 m^2} \left( \frac{\partial^2}{\partial a^2} -
   \frac{\langle \tilde{\chi} | \frac{\partial^2}
{\partial a^2} | \tilde{\chi} \rangle }
{\langle \tilde{\chi} | \tilde{\chi} \rangle}
 \right) \tilde{\chi}
  \label{schtilde}
  \ee
  
  Indeed, on substituing $\tilde{\psi}_K$ and $\tilde{\chi}_K$  for
  $\tilde{\psi}$ and $\tilde{\chi}$ into eqs. (\ref{hjtilde}) and
  (\ref{schtilde}) and retaining terms to ${\cal O}(m^2)$,
  ${\cal O}(m^0)$, ${\cal O}(m^{-2})$, one obtains from (\ref{hjtilde}): 
  
  \be
  - \frac{1}{2} S_0'^2 + V =0
  \label{hjmdue}
  \ee 
  
  \be
  - \hbar i \frac{N_K' S_0'}{N_K} + \frac{\hbar i S_0''}{2} +
  \langle \hat{H}^M \rangle_0 =0
  \label{hjmzero}
  \ee
  
  \be
  \frac{\hbar^2}{2} \left( 2 \frac{N_K'^2}{N_K^2} 
  - \frac{2 S_0' \sigma_2'}{\hbar^2}
  - \frac{N_K''}{N_K} \right) +   
  \langle \hat{H}^M \rangle_{-2} = - \frac{\hbar^2}{2}  \langle \bar{\De}^2 
  \rangle_0
  \label{hjmmendue}
  \ee
  
  \noindent 
  where a prime denotes a derivative with respect to $a$. Eqs. (\ref{hjmdue}),
  (\ref{hjmzero}) and (\ref{hjmmendue}) are respectively the equations
  ${\cal O}(m^2)$, ${\cal O}(m^0)$, ${\cal O}(m^{-2})$ and by 
  $\left[ \langle \hat{H}^M \rangle_0 \,, \,\,\; 
  \langle \bar{\De}^2 \rangle_0 \right]$ and
  $\langle \hat{H}^M \rangle_{-2}$ we mean the corresponding terms of 
  ${\cal O}(m^0)$ and ${\cal O}(m^{-2})$ respectively. Analogously from eq.
  (\ref{schtilde}) one obtains:
  
  \be
  H_0^M -  \langle \hat{H}^M \rangle_0 + \hbar i S_0' \left(
  \frac{N_K'}{N_K} + \frac{i S_1'}{\hbar} \right) = 0
  \label{schmzero}
  \ee
  
  \begin{eqnarray} \lefteqn{
  \left( H_{-2}^M - \langle \hat{H}^M \rangle_{-2}  \right)
  - \hbar^2 \frac{N_K'}{N_K} \left( \frac{N_K'}{N_K} + \frac{i S_1'}{\hbar}
  \right) - S_0' \eta_2' =
  } \nonumber \\ & & \frac{\hbar^2}{2} \left(
   \langle \bar{\De}^2 \rangle_0 - \frac{N_K''}{N_K} - i \frac{S_1''}{\hbar} +
  \frac{ S_1'^2}{\hbar^2} - 2 \frac{i S_1'}{\hbar} \frac{N_K'}{N_K}\right)
  \label{schmmendue}
  \end{eqnarray}
  
  \noindent 
  to ${\cal O}(m^0)$ and ${\cal O}(m^{-2})$ respectively (where we 
have defined the c-number $H^M$ by $\hat{H}^M \chi_K = H^M \chi_K$).
%
On comparing the two
  ${\cal O}(m^0)$ equations and eliminating  $\langle \hat{H}^M \rangle_0$
  one obtains:
  
  \be
  H_0^M + \frac{i \hbar S_0''}{2} - S_0' S_1' =0
  \label{mzeromzero}
  \ee
  
  \noindent 
  which agrees with the corresponding equation obtained on 
substituing for $\Psi$
  directly in the WDW equation (\ref{wdw}) \cite{kiefer1}.
  Similarly on composing the two
  ${\cal O}(m^{-2})$  equations one can eliminate 
  $\left( \langle \hat{H}^M \rangle_{-2} + \frac{\hbar^2}{2} 
  \langle \bar{D}^2 \rangle_{0} \right) $ obtaining:
  
  \be
  - S_0' \eta_2' + H_{-2}^M - S_0' \sigma_2' + i \frac{S_1''}{2} \hbar
  - \frac{S_1'^2}{2} = 0
\label{mmduemmdue}
  \ee
  
  \noindent 
  again in agreement with the corresponding expression obtained directly from
  the WDW equation \cite{kiefer1}. Thus eqs. (\ref{mzeromzero}) 
and (\ref{mmduemmdue}) confirm our identification
of $\tilde{\psi}_k$ and $\tilde{\chi}_k$ with $\tilde{\psi}$ and $\tilde{\chi}$
respectively.

The above equations (\ref{schmzero}) and (\ref{schmmendue}) then 
lead to the following expression for eq. (\ref{sch1}):

\begin{eqnarray} \lefteqn{
\left\{ \left( H_0^M + \frac{1}{m^2} H_{-2}^M \right) - i \hbar 
\frac{\partial}{\partial \eta} \right\} 
e^{- \frac{i}{\hbar} \int^\eta \left( \langle \hat{H}^M \rangle_0
- \frac{1}{m^2} \langle \hat{H}^M \rangle_{-2} \right) \de \eta'}
\tilde{\chi}_K = 
} \nonumber \\ & &
\frac{\hbar^2}{m^2} \frac{N_K'}{N_K} 
\left( \frac{N_K'}{N_K} + \frac{i S_1'}{\hbar} \right)
\tilde{\chi}_K + 
\nonumber \\ & &
\frac{\hbar^2}{2 m^2} 
\left( \frac{\langle \tilde{\chi} | \frac{\partial^2}
{\partial a^2} | \tilde{\chi} \rangle_0 }
{\langle \tilde{\chi} | \tilde{\chi} \rangle}
- \frac{N_K''}{N_K} - i \frac{S_1''}{\hbar} +
  \frac{ S_1'^2}{\hbar^2} - 2 \frac{i S_1'}{\hbar} \frac{N_K'}{N_K}\right)
\tilde{\chi}_K
\label{chiks}
\end{eqnarray}
\noindent where we now have $\chi_{K s} \equiv e^{- \frac{i}{\hbar} 
\int^\eta \left( \langle \hat{H}^M \rangle_0
- \frac{1}{m^2} \langle \hat{H}^M \rangle_{-2} \right) \de \eta'}
\tilde{\chi}_K $.
Note that $\langle \chi| \bar{\De}| \chi \rangle$ $=$
$\langle \tilde{\chi}| \frac{\partial}{\partial a}| \tilde{\chi} \rangle$ 
$=$ $0$
now becomes (see also eq. (\ref{schmzero})):

\be
\langle \tilde{\chi}| \frac{\partial}{\partial a}| \tilde{\chi} \rangle =
\int \de \phi \left[ \chi_K^* \left( \frac{N_k'}{N_k} + i 
\frac{S_1'}{\hbar} + i \frac{\eta_2'}{m^2\hbar} \right) \chi_K \right] = 0 \,.
\label{chitdachit}
\ee

  One may now check, as in eq. (\ref{unit1}), whether a violation of unitarity
  occurs, obtaining on using eqs. (\ref{chiks}) and (\ref{chitdachit}):
  
  \be
  i \hbar \frac{\partial}{\partial \eta} \int {\chi}_{K s}^* {\chi}_{K s}
  \de \phi = i \hbar S_0' \int \left[ \tilde{\chi}_K^* \frac{\partial}{\partial a}
  \tilde{\chi}_K + c.c \right] \de \phi = 0
  \ee
  
  \noindent in agreement with our general result eq. (\ref{unit1}).
As before the result is a consequence of the presence of back-reaction terms 
$\langle \hat{H}^M \rangle$ and 
$\frac{\hbar^2}{2 m^2} \frac{\langle \tilde{\chi} | \frac{\partial^2}
{\partial a^2} | \tilde{\chi} \rangle }
{\langle \tilde{\chi} | \tilde{\chi} \rangle}$, indeed as can be seen from 
eq. (\ref{chiks}) such terms are subtracted from the corresponding 
operators acting on $\tilde{\chi}_K$, thus leading to zero on integrating 
over all matter configurations. The presence of back-reaction terms also 
modifies the definitions of $N_K$ and $\sigma_2$ with respect to those 
employed elsewhere \cite{kiefer1}, indeed from eq. (\ref{hjmzero}) and 
(\ref{hjmmendue}) one sees that these quantities are not associated 
with pure gravity, but include the mean effect of matter (see eq. 
(\ref{hjtilde})).

  
One may alternatively attempt to use instead of $N_K$ a prefactor $N_G$ 
satisfying
  eq. (\ref{hjmzero}) for $\langle \hat{H}^M \rangle_0 =0$ (pure gravity)
and introduce simultaneously
an adiabatic phase factor in eq. (\ref{psichikief}),  
 that is using instead of $\tilde{\psi}_K$:
  
  \be
  \psi_K = e^{-i \int^a A \de a'} \tilde{\psi}_K
  \ee
  
  \noindent which will modify eq. (\ref{hjmzero}) leading to 
\cite{kiefer2,datta}:
  
  \be
  - \hbar i S_0' \left( \frac{N_G'}{N_G} - \langle \frac{\partial}{\partial a}
  \rangle_0 \right) + \frac{\hbar i S_0''}{2} + \langle \hat{H}^M \rangle_0 =
  i \hbar S_0' \langle \frac{\partial}{\partial a}
  \rangle_0 +  \langle \hat{H}^M \rangle_0 = 0
  \ee
  
  \noindent
  which however is in disagreement with the distinction between the adiabatic
  and dynamical phases \cite{venturi,wudka}. Indeed as a consequence of 
  reparametrization invariance the dynamical
  phases of gravity and matter wave-functions cancel (Einstein equation). 
  Similarly it is known
  that the light and heavy systems have equal and opposite adiabatic phases
  \cite{mead,venturi}. Further the presence of the adiabatic phase and the 
fluctuations on the right hand sides of eqs. (\ref{eqpsi}) and (\ref{eqchi}) 
are associated with corrections to the 
adiabatic approximation, that is they are 
related to transitions between different states and are expected to be small
\cite{casadio} 
(if such were not the case one would not have the usual evolution equations).
 
  Let us now compare our result with yet another approach \cite{kim}. 
  Our equation 
  (\ref{eqpsi}) was obtained by substituing the BO decomposition into
  the WDW equation and contracting with $\chi^*$. One may also expand
  $|\chi \rangle$ on a suitable orthonormal basis $|l \rangle$:
  
  \be
  |\chi \rangle  = \sum_l c_l(\eta) | l \rangle
  \label{expans}
  \ee
  
  \noindent 
  and contract with respect to $\langle n |$. One obtains \cite{kim} 
  instead of eq.
  (\ref{eqpsi}):
  
  \be
  \frac{\hbar^2}{2m^2} \De^2_n \psi + m^2 V + \frac
  {\langle n | \hat{H}^M | \chi \rangle}{\langle n | \chi \rangle}
  = - \frac{\hbar^2}{2 m^2} 
  \frac{\langle n | \bar{\De}^2_n | \chi \rangle}{\langle n | \chi \rangle} \psi 
  \label{psikif}
  \ee
  
  \noindent where:
  
  \be
  \De_n \equiv \frac{\partial}{\partial a} + i A_n \,; \;\;\;\;\;
  \bar{\De}_n \equiv \frac{\partial}{\partial a} - i A_n
  \ee
  
  \noindent with:
  
  \be
  A_n \equiv - i \frac{\langle n | \frac{\partial}{\partial a} |\chi \rangle} 
  {\langle n | \chi \rangle} 
  \ee
  
  \noindent 
  Of course if one multiplies eq. (\ref{psikif}) by $c_n^*$ and sums over $n$
  eq. (\ref{eqpsi}) is again obtained.
  
  As before, one may now multiply eq. (\ref{psikif}) by $\chi$ and subtract
  it from eq. (\ref{wdw}) obtaining:
  
  \be
  \psi \left( \hat{H}^M - \frac{ \langle n | \hat{H}^M | \chi \rangle}
  {\langle n| \chi \rangle}
   \right) \chi +
  \frac{\hbar^2}{m^2} \left( \De_n \psi \right) \bar{\De}_n \chi =
  - \frac{\hbar^2}{m^2} \psi \left( \bar{\De}_n^2 - 
  \frac{ \langle n | \bar{\De}^2 | \chi \rangle}{\langle n | \chi \rangle} \right)
  \chi 
  \ee
  
  \noindent 
  and on multiplying the above by $c_n^*$ and summing over $n$ 
  the usual result eq. (\ref{eqchi}) is obtained. Thus, again, we have 
  separated the gravitational and matter equations of motion and
  equations (\ref{hjtilde}) and (\ref{schtilde}) now become:
  
  \be
  \left( \frac{\hbar^2}{2 m^2} \frac{\partial^2}{\partial a^2} +
  m^2 V + \frac{\langle n| \hat{H}^M |\chi \rangle}
  {\langle n|\chi\rangle} \right) \tilde{\psi}_n =
  - \frac{\hbar^2}{2 m^2} \frac{\langle n| \bar{\De}_n^2 |\chi \rangle} 
  {\langle n|\chi\rangle} \tilde{\psi}_n
  \label{hjtildekif}
  \ee
  
  \noindent and:
  
  \be
  \left( \hat{H}^M - \frac{\langle n| \hat{H}^M |\chi \rangle}
  {\langle n|\chi\rangle}  \right) \tilde{\chi}_n
  +  \frac{\hbar^2}{m^2} \frac{\partial \ln \tilde{\psi}_n}{\partial a}
  \frac{\partial \tilde{\chi}_n}{\partial a} =
  - \frac{\hbar^2}{2 m^2} \left( \frac{\partial^2}{\partial a^2} -
  \frac{\langle n| \bar{\De}_n^2 | \chi \rangle}{\langle n|\chi\rangle} 
   \right) \tilde{\chi}_n
  \ee
  
  \noindent where we have defined:
  
  \be
  \tilde{\psi}_n \equiv e^{i \int^a A_n \de a'} \psi \equiv
  e^{i \int^a \left( A_n - A \right) \de a'} \tilde{\psi}
  \ee
  
  \be
  \tilde{\chi}_n \equiv e^{-i \int^a A_n \de a'} \chi \equiv
  e^{-i \int^a \left( A_n - A \right) \de a'} \tilde{\chi}
  \ee
  
  The above expressions simplify considerably if $| \chi \rangle = 
  | n \rangle$ with $\chi_n = \langle \phi | n \rangle$, in which case one has:
  
  \be
  \left( \frac{\hbar^2}{2 m^2} \frac{\partial^2}{\partial a^2} +
  m^2 V + \langle n| \hat{H}^M | n \rangle
  \right) \tilde{\psi}_{nn} =
  - \frac{\hbar^2}{2 m^2} \langle n| \bar{\De}_{nn}^2 | n \rangle 
  \tilde{\psi}_{nn}
  \label{hjtildekif2}
  \ee
  
  \noindent and:
  
  \be
  \left( \hat{H}^M - \langle n| \hat{H}^M | n \rangle
  \right) \tilde{\chi}_{nn}
  +  \frac{\hbar^2}{m^2} \frac{\partial \ln \tilde{\psi}_{nn}}{\partial a}
  \frac{\partial \tilde{\chi}_{nn}}{\partial a} =
  - \frac{\hbar^2}{2 m^2} \left( \frac{\partial^2}{\partial a^2} -
  \langle n| \bar{\De}_{nn}^2 | n \rangle 
   \right) \tilde{\chi}_{nn}
  \label{schtildekif}
  \ee
  
  \noindent where we have correspondingly defined:
  
  \be
  \De_{nn} \equiv \frac{\partial}{\partial a} + i A_{nn} \,; \;\;\;\;\;
  \bar{\De}_{nn} \equiv \frac{\partial}{\partial a} - i A_{nn}
  \ee
  
  \noindent with:
  
  \be
  A_{nn} \equiv - i \langle n | \frac{\partial}{\partial a} | n \rangle =
  i {\langle n |  \frac{\stackrel{\leftarrow}{\partial}}{\partial_a}  
  | n \rangle} 
  \ee
  
  \noindent and:
  
  \be
  \tilde{\psi}_{nn} = e^{i \int^a A_{nn} \de a'} \psi_n 
  \ee
  
  \be
  \tilde{\chi}_{nn} = e^{-i \int^a A_{nn} \de a'} \chi_n 
  \ee
  
  \noindent Similarly, instead of eq. (\ref{chisch}) one obtains:
  
  \be
  \chi_{ns} \equiv e^{- \frac{i}{\hbar} \int^\eta \langle n| \hat{H}^M 
  |n \rangle \de\eta'
  - i \int^{a_n(\eta)} \de a' A_{nn}} \chi_n 
  \label{chischn}
  \ee
  
  \noindent where in this case $a_n(\eta)$ is the trajectory
  obtained in the semiclassical limit
  for $\tilde{\psi}_{nn}$.
  
  We now note that the r.h.s. of eq. (\ref{sch1}) (similarly for eq.
  (\ref{schtildekif})) always consists of an operator acting on a state
  minus its expectation value with respect to that state, that is it
  is associated with fluctuations. 
  Let us then denote by $| \nu \rangle \left( \chi_\nu \right)$ the
  eigenstates obtained instead of $| n \rangle \left( \chi_n \right)$ 
  on omitting fluctuations in eq. (\ref{schtildekif}). Correspondingly
  one will have a solution $\chi_{\nu s}$ to the Schr\"{o}dinger equation:
    
  \begin{eqnarray} &
  \left( \hat{H}^M - i \hbar \frac{\partial}{\partial\eta} \right)
  e^{ - \frac{i}{\hbar} \int^\eta 
\langle \nu |\hat{H}^M | \nu \rangle \de \eta' -
  i \int^{a_\nu(\eta)} \de a' A_{\nu\nu}} \chi_\nu \equiv 
   \nonumber \\ &  
  \left( \hat{H}^M - i \hbar \frac{\partial}{\partial\eta} \right)
  \chi_{\nu s} = 0
  \label{schnu}
  \end{eqnarray}
  
  \noindent
  From eq. (\ref{schnu}) one can obtain some information about the
  orthonormal basis $| \nu \rangle$, indeed on contracting the above
  equation with $\chi_\lambda^* \left( \langle \lambda | \right)$
  one sees that it is identically satisfied for for
  $\langle \lambda | = \langle \nu | $ but for
  $\langle \lambda | \neq \langle \nu | $ one has:
  
  \be
  \langle \lambda | \left( \hat{H}^M - i \hbar \frac{\partial}{\partial\eta} 
  \right) | \nu \rangle = 0
  \label{lamnu}
  \ee
  
  \noindent
  The above equations (\ref{schnu}) and (\ref{lamnu}) allow us to identify
  the $| \lambda \rangle$ with the eigenstates of the time-dependent 
  invariants \cite{lewis} 
  associated with our time dipendent matter hamiltonian $\hat{H}^M$.
  
  On using eq. (\ref{chisch}), in analogy with eq. (\ref{expans}), 
  one may express 
  $\chi_s$ in our new basis:
  
  \begin{eqnarray} \lefteqn{
  \chi_s = e^{- \frac{i}{\hbar} \int^\eta \langle \hat{H}^M \rangle \de\eta'
  - i \int^{a(\eta)} \de a' A} \sum_\lambda c_\lambda(\eta) \chi_\lambda =
  } \nonumber \\ & 
  e^{- \frac{i}{\hbar} \int^\eta \langle \hat{H}^M \rangle \de\eta'
  - i \int^{a(\eta)} \de a' A} 
  \sum_\lambda c_\lambda(\eta) 
  e^{\frac{i}{\hbar} \int^\eta \langle \lambda | \hat{H}^M | \lambda \rangle 
  \de\eta' + i \int^{a_\lambda(\eta)} \de a' A_{\lambda\lambda}} \chi_{\lambda s}
  \label{chischlam}
  \end{eqnarray}
  
  \noindent which on substituing in eq. (\ref{sch1}) leads to:
  
  \be
  \sum_\lambda \left( - \langle \hat{H}^M \rangle - \hbar \dot{a} A +
  \langle \lambda | \hat{H}^M | \lambda \rangle + \hbar \dot{a}_\lambda(\eta) 
  A_{\lambda\lambda} - i \hbar
  \frac{\dot{c}_\lambda}{c_\lambda} \right) \chi_\lambda \propto 
  {\rm fluctuations}
  \ee
  
  \noindent and on neglecting fluctuations (and denoting the quantities 
  then obtained by the additional superscript $0$) one has:
  
  \be
  c_\lambda^0(\eta) = c_\lambda^0(0) e^{
   \frac{i}{\hbar} \int_0^\eta \langle \hat{H}^M \rangle^0 \de\eta'
  + i \int^{a_0(\eta)} \de a' A^0
  - \frac{i}{\hbar} \int_0^\eta \langle \lambda | \hat{H}^M | \lambda \rangle 
  \de\eta' - i \int^{a_\lambda} \de a' A_{\lambda\lambda}}
  \ee
  
  \noindent
  From the above one then obtains:

  \be
  \chi_s^0 = \sum_\lambda c_\lambda^0(0) e^{
  - \frac{i}{\hbar} \int^\eta \langle \lambda | \hat{H}^M | \lambda \rangle 
  \de\eta' - i \int^{a_\lambda} \de a' A_{\lambda\lambda}} \chi_\lambda =
  \sum_\lambda c_\lambda^0(0) \chi_{\lambda s}
  \ee
  
  \noindent
  which is the form for the general solution of the Schr\"{o}dinger equation
  for a time dependent Hamiltonian in terms of the eigenfunctions of the 
  time-dependent invariants \cite{lewis}.
  
  If we do not neglect the fluctuations we may write:
  
  \begin{eqnarray} &
  \chi_s = \sum_\lambda \left[ c_\lambda^0(0) + \delta_\lambda(\eta) \right]
  e^{
  - \frac{i}{\hbar} \int^\eta \langle \lambda | \hat{H}^M | \lambda \rangle 
  \de\eta' - i \int^{a_\lambda} \de a' A_{\lambda\lambda}} \chi_\lambda 
   \nonumber \\  &
  = \sum_\lambda \left( c_\lambda^0(0) + \delta_\lambda(\eta) \right)  
  \chi_{\lambda s}
  \end{eqnarray}
  
  \noindent
  where the $\delta_\lambda(\eta)$ will be determined by the fluctuations and,
  on substituting into our general probability conservation constraint eq.
  (\ref{unit1}), we obtain:
  
  \begin{eqnarray} \lefteqn{
  i \hbar \frac{\partial}{\partial \eta} \int \chi_s^* \chi_s \de \phi =
  i \hbar \frac{\partial}{\partial \eta} \sum_\lambda \left|
  c_\lambda^0(0) + \delta_\lambda(\eta) \right|^2 =
  } \nonumber \\ & &
  i \hbar \sum_\lambda \left[ c_\lambda^{0*}(0) \dot{\delta}_\lambda(\eta)  +
  {\dot{\delta}_\lambda^*} (\eta) c_\lambda^0(0) + {\dot{\delta}_\lambda^*}(\eta) 
  \delta_\lambda(\eta) +\delta_\lambda^*(\eta) \dot{\delta}_\lambda(\eta)  
  \right] =0
  \end{eqnarray}
  
  \noindent
  which just reflects the fact that an increase of the weight $\delta_\lambda
  (\eta)$ of one state is compensated by the decrease of another.
  Thus there is no overall violation of unitarity, it is only
  on considering a single state that the evolution appears to be non unitary
  \cite{wudka}, and the result is obtained without having to resort to the
limit $m^2 \rightarrow \infty $\cite{kim}.
  
Let us conclude: the approach we have illustrated consists of writing
the total matter-gravity wave function $\Psi$ in a factorized form
$\psi \chi$ (Born-Oppenheimer) satisfying coupled equations, where
$\psi$ describes gravitation with the back-reaction of matter
and $\chi$ describes matter on a curved background with both equations
including fluctuations.
It can be seen, on neglecting fluctuations and considering a semiclassical
approximation for gravity, that the two wave functions $\psi$ and $\chi$
can be interpreted as describing gravitation with a back-reaction 
due to the mean
energy of matter and matter satisfying a Schwinger-Tomonaga equation
and following gravitation adiabatically. The initial coupled equations
themselves are a consequence of the factorization ansatz and do not
involve any approximation.

Other approaches in general also write a factorized wave function which is
substitued in the WDW equation, however they do not proceed to obtain
coupled equations for matter and gravitation, but the diverse
terms in their resulting single equation are identified
by equating powers of $m^2$ and subsequentely a non-unitary evolution
for matter is obtained. We have seen in one case \cite{kiefer1}
that if such an expansion is performed in our coupled equations which
include the effects of all backreaction
no difficulty arises with unitary evolution and in the other
case \cite{kim} that there is no need to neglect lower powers of
$m^2$ in order to obtain unitary evolution.


\begin{thebibliography}{99}
  
  \bibitem{mead} See for example 
  
  C. A. Mead and D. G. Truhlar, {\em J. Chem. Phys.} {\bf 70}, 2284 (1979)
  
  C. A. Mead, {\em J. Chem. Phys.} {\bf 49}, 2333 (1984)
  
  \bibitem{brout} R. Brout, {\em Found. Phys.} {\bf 17}, 603 (1987)
  
  R. Brout and G. Venturi, {\em Phys. Rev.} {\bf D15}, 2439 (1989)
  
  \bibitem{banks} T. Banks, {\em Nucl. Phys.} {\bf B 249}, 332 (1985)
  
  \bibitem{kiefer1} C. Kiefer and T. P. Singh 
  {\em Phys. Rev.} {\bf D 44}, 1067 (1991)
  
  \bibitem{kiefer2} C. Kiefer, {\em Class. Quantum Grav.} {\bf 9}, 147 (1992)
  
  C. Kiefer, in {\em Canonical Gravity - From Classical to Quantum},
   eds. J. Ehlers and H. Friedrich (Springer, Berlin, 1994).
  
  \bibitem{kim} S. P. Kim, {\em Phys. Lett.} {\bf A 205}, 359 (1995)
  
  S. P. Kim, {\em Phys. Rev.} {\bf D 52}, 3382 (1995)
  
  \bibitem{coscos} For the case of just a cosmological constant $\Lambda$
  one has $2 V = - a^2 + \frac{\Lambda}{3} a^4$.
  
  \bibitem{dewitt} B. S. DeWitt, {\em Phys. Rev.} {\bf 160}, 113 (1967)
  
  \bibitem{norsta} The diverse expressions may be written in a more symmetric
  fashion by considering normalized states $| \chi \rangle_N =
  \frac{| \chi \rangle}{\left( \langle \chi | \chi \rangle \right)^{\frac{1}{2}}}$
  in which case $A = - \frac{i}{2} \left[ _N\langle \chi | 
  \frac{\partial}{\partial a} |\chi
  \rangle_N - _N\langle \chi | \frac{\stackrel{\leftarrow}{\partial}}{\partial a}
  |\chi \rangle_N \right] = - \frac{i}{2} _N\langle \chi | 
  \frac{\stackrel{\leftrightarrow}{\partial}}{\partial a}|\chi
  \rangle_N $ which is explicitily Hermitian.
  
  \bibitem{clalim}Of course we assume that a classical limit exists which
  implies that $\left| \psi \right|^2$ is strongly peaked on the classical
  trajectory $a(\eta)$. For a molecule this will correspond to considering
  the motion of the nuclei to be quasi-classical while that of the electrons 
  is quantum-mechanical. Clearly the semiclassical limit is an addition
to the Born-Oppenheimer factorization and is necessary for time to emerge.
In the semiclassical limit, in a path integral representation for the
wave function, neighbouring paths will tend to yield cancelling
contributions on account of the rapid variation of the phase associated
the exponential of the (effective) action. An exception to this rule
occurs at stationary points of the exponent and the associated paths are
related to classical trajectories. It is clear that for this limit
to exist the fluctuations about the solutions to the classical equations
of motions must be small and the integral over them finite. In general
this leads to a constraint on the effective potential associated with the
fluctuations, should this not be satisfied one could have for example
fluctuations which increase exponentially in time which, of course, signal
an instability.
   
  \bibitem{phases} Let us note the distinction in the wave function between
  the adiabatically induced phase (related to $A$) and the dynamical
  phase (related to $\langle \hat{H}^M \rangle$).
  
  \bibitem{venturi1} G. Venturi, {\em Astroparticle Physics}, {\bf 1}, 417 (1993)
  
  \bibitem{datta} D. P. Datta, {\em Phys.
  Rev.} {\bf D 40}, 574 (1993) and {General Rel. and Grav.} {\bf 27},
  341 (1995)
  
  \bibitem{venturi} G. Venturi, {\em Class. and Quantum Grav.} {\bf 7},
  1075 (1990)

\bibitem{casadio} See for example R. Casadio and G. Venturi
{\em Class. and Quantum Grav.} {\bf 12}, 1267 (1995)
  
  \bibitem{wudka} J. Wudka, {\em Phys. Rev.} {\bf D 41}, 712 (1990)
  
  J. Vidal and J. Wudka, {\em Phys. Rev.} {\bf A 44}, 5383 (1991)

  
  \bibitem{lewis} That is eigenstates of explicitly time dependent
  non-trivial Hermitian operators $\hat{I}(\eta)$ satisfying
  $\frac{\de \hat{I}}{\de \eta} \equiv \frac{\partial \hat{I}}{\partial \eta} -
  \frac{i}{\hbar} \left[ \hat{I}, \hat{H} \right] =0$. See
  H. R. Lewis jr. and W. B. Riesenfeld, {\em
  Journal of Math. Phys.} {\bf 10}, 1458 (1969)
  
  \end{thebibliography}
  \end{document}